\documentclass[12pt]{article}

\usepackage{color}
\definecolor{darkblue}{rgb}{0.1,0.1,.7}
\usepackage[colorlinks, linkcolor=darkblue, citecolor=darkblue, urlcolor=darkblue, linktocpage]{hyperref}
\usepackage[]{amsmath}
\usepackage[]{graphicx}
\usepackage[]{latexsym}
\usepackage[utf8]{inputenc}
\usepackage{slashed,graphicx,color,amsmath,amssymb}
\usepackage[mathscr]{eucal}
\usepackage{mathrsfs}
\usepackage[margin=10pt,font=small,labelfont=bf]{caption}
\usepackage{amscd}
\usepackage{bm}
\usepackage{xcolor}
\usepackage{upgreek}
\usepackage[square, comma, sort&compress,numbers]{natbib}
\usepackage[all,cmtip]{xy}
\usepackage[margin=1.7in]{geometry}
\usepackage[color=cyan!30!white,linecolor=red,textsize=footnotesize]{todonotes}
\numberwithin{equation}{section}
\hypersetup{
	pdfstartview={FitH},    
	pdftitle={Modular invariance and uniqueness of TT deformed CFT},    
	pdfauthor={Aharony, Datta, Giveon, Jiang, Kutasov},     
	colorlinks=true,       
	linkcolor=blue,          
	citecolor=blue!59!black! ,        
	filecolor=magenta,      
	urlcolor=blue           
}

\newcommand{\tr}{\mathrm{Tr}\,}

\newcommand{\rE}{\mathrm{E}}

\def\ttb{T{\bar{T}}}
\def\btau{{\bar{\tau}}}

\def\pd{\partial}

 \makeatletter
 \g@addto@macro\bfseries{\boldmath}
 \makeatother

\begin{document}
\vspace*{-.6in} \thispagestyle{empty}
\vspace{.2in} {\Large
\begin{center}
{\bf Modular invariance and uniqueness \\ of $T\bar T$ deformed CFT}
\end{center}}
\vspace{.2in}
\begin{center}
{Ofer Aharony$^{1}$, Shouvik Datta$^2$, Amit Giveon$^3$, Yunfeng Jiang$^2$ \& David Kutasov$^4$}
\\
\vspace{.3in}
\small{
$^1$  \textit{Department of Particle Physics and Astrophysics, \\ Weizmann Institute of Science, \\
Rehovot 7610001, Israel.}\\  \vspace{.3cm}
$^2$\textit{Institut f{\"u}r Theoretische Physik,
ETH Z{\"u}rich}},\\
\small{\textit{Wolfgang Pauli Strasse 27,
CH-8093 Z{\"u}rich, Switzerland.}
}
\\ \vspace{.3cm}
$^3$\textit{Racah Institute of Physics, The Hebrew University, \\
Jerusalem 91904, Israel.
}
\\ \vspace{.3cm}
$^4$\textit{EFI and Department of Physics, University of Chicago, \\
5640 S.~Ellis Av., Chicago, IL 60637, USA. \\
}
\vspace{.1in}

\end{center}

\vspace{.2in}

\begin{abstract}
\normalsize{Any two dimensional quantum field theory that can be consistently defined on a torus is invariant under modular transformations. In this paper we study families of quantum field theories labeled by a dimensionful parameter $t$, that have the additional property that the energy of a state at finite $t$ is a function only of $t$ and of the energy and momentum of the corresponding state at $t=0$, where the theory becomes conformal. We show that under this requirement, the partition sum of the theory at $t=0$ uniquely determines the partition sum (and thus the spectrum) of the perturbed theory, to all orders in $t$, to be that of a $T\bar T$ deformed CFT. Non-perturbatively, we find that for one sign of $t$ (for which the energies are real) the partition sum is uniquely determined, while for the other sign we find non-perturbative ambiguities. We characterize these ambiguities and comment on their possible relations to holography.
}

\end{abstract}

\vskip 1cm \hspace{0.7cm}

\newpage

\setcounter{page}{1}

\noindent\rule{\textwidth}{.1pt}\vspace{-1.2cm}
\begingroup
\hypersetup{linkcolor=black}
\tableofcontents
\endgroup
\noindent\rule{\textwidth}{.2pt}

\section{Introduction}
\label{sec:1}

A $T\bar T$ deformed conformal field theory (CFT), a non-local theory that has recently received some attention \cite{Zamolodchikov:2004ce,Smirnov:2016lqw,Cavaglia:2016oda,McGough:2016lol,Giveon:2017nie,Dubovsky:2017cnj,Giveon:2017myj,Shyam:2017znq,Asrat:2017tzd,Giribet:2017imm,Kraus:2018xrn,Cardy:2018sdv,Cottrell:2018skz,Aharony:2018vux,Dubovsky:2018bmo,Taylor:2018xcy,Bonelli:2018kik,Datta:2018thy,Donnelly:2018bef,Babaro:2018cmq,Conti:2018jho,Chakraborty:2018kpr,Chen:2018eqk}, is obtained by adding to the Lagrangian of a two dimensional CFT an irrelevant operator bilinear in stress tensors in a specific manner. The corresponding coupling, $t$, has holomorphic and anti-holomorphic dimensions $(-1,-1)$ (i.e.~it scales like length squared). Despite the fact that the perturbation is irrelevant, and thus corresponds to a flow up the renormalization group, the authors of \cite{Smirnov:2016lqw,Cavaglia:2016oda} showed that the resulting theory is in some sense solvable. In particular, they computed the spectrum of the theory on a circle of radius $R$.

In \cite{Datta:2018thy} it was shown that the spectrum found in \cite{Smirnov:2016lqw,Cavaglia:2016oda}  leads to a modular invariant torus partition sum. In terms of the dimensionless coupling,
$\lambda \sim {t}/{R^2}$,
which can be thought of as the value of the coupling $t$ at the scale $R$, it was found in \cite{Datta:2018thy}  that the partition sum satisfies
\begin{align}
\label{eq:modularZdef}
\mathcal{Z}\left(\left.\frac{a\tau+b}{c\tau+d},\frac{a\bar{\tau}+b}{c\bar{\tau}+d}\right|\frac{\lambda}{|c\tau+d|^2}\right)
=\mathcal{Z}(\tau,\bar{\tau}|\lambda),
\end{align}
where $\tau$ is the modular parameter of the torus, $a,b,c,d\in\mathbb{Z}$ and $ad-bc=1$. At $\lambda=0$, (\ref{eq:modularZdef}) reduces to the modular invariance of the original CFT. In general, $\lambda$ transforms as a modular form of weight $(-1,-1)$.

In this paper we revisit the torus partition sum of $T\bar T$ deformed CFTs from a different perspective. The starting point of our discussion is the observation that modular invariance of the partition sum, (\ref{eq:modularZdef}), is guaranteed on general grounds to be a property of any theory that can be consistently formulated on a torus, since modular transformations correspond to reparametrizations of the torus.\footnote{More precisely, modular invariance leads to equation  (\ref{eq:modularZdef}) when the theory has a single dimensionful parameter of dimension $(-1,-1)$. The generalization to cases where the parameter has a different dimension is straightforward.}

For example, on a rectangular torus of size $L_1\times L_2$, the torus partition sum must obey $Z(L_1,L_2, t)=Z(L_2,L_1,t)$, due to the freedom of relabeling the axes. At the same time, if we view $L_1$ as the circumference of the circle on which the theory lives, $L_1=2\pi R$, and $L_2$ as the inverse temperature, $L_2=\beta$, the transformation $L_1\leftrightarrow L_2$ acts non-trivially on $R$ and on $\tau_2=\beta/2\pi R$. In terms of the dimensionless coupling $\lambda$, this leads to (a special case of) (\ref{eq:modularZdef}).


$T\bar T$ deformed CFTs have the additional property that the energies of states in the deformed theory depend in a universal way on
the energies and momenta of the corresponding states in the undeformed theory. In particular, this deformation does not lift any degeneracies in the undeformed spectrum. Note that this is a highly restrictive property, and it is natural to ask whether the class of theories that have this property (and the other properties mentioned above — a single scale, and modular invariance) is larger. For example, one can ask whether this class includes any of the other irrelevant deformations discussed in \cite{Smirnov:2016lqw}.

We show that the answer is negative -- under the above assumptions, the torus partition sum (and thus the spectrum) of the deformed theory is uniquely determined in terms of that of the undeformed theory to all orders in $\lambda$. Of course, the resulting partition sum and spectrum must then agree with those of a $T\bar T$ deformed CFT, and we show that this is indeed the case.

Note that the assumption about the energies of states stated above only applies to states that have a smooth $\lambda\to 0$ limit, i.e.~states whose contributions to the partition sum have a perturbative expansion in $\lambda$. The deformed theory could have additional states whose energies diverge in the limit $\lambda\to 0$; in general, these give rise to non-perturbative contributions to the partition sum $\mathcal{Z}(\tau,\bar{\tau}|\lambda)$, which need to be discussed separately. One may think of these states as giving different high-energy completions for the deformed theory at finite $t$.

Note also that the assumption that the theory has a single scale, associated with the coupling $t$, is quite non-trivial. In general, theories with irrelevant couplings develop an infinite number of scales, associated with the coefficients of all possible irrelevant operators consistent with the symmetries. This is another way of saying that such theories are not renormalizable. When viewed as effective field theories, they also depend on the choice of UV cutoff. We are assuming that in our case, there is only one scale, i.e. that the coefficients of all operators other than the one that couples to $t$ can be consistently set to zero, and that there is no dependence on the UV cutoff.

The all orders partition sum satisfies a first order differential equation in the dimensionless coupling $\lambda$. To study the theory non-perturbatively in $\lambda$, we assume\footnote{For the $T\bar T$ case, this assumption follows from the results of \cite{Cardy:2018sdv,Dubovsky:2018bmo}, but here we take a broader point of view.} that this equation persists beyond perturbation theory, and analyze its solutions. We find that for $\lambda>0$ (the sign considered in \cite{Smirnov:2016lqw,Cavaglia:2016oda,Giveon:2017nie}, for which the spectrum of energies is real for small $\lambda$), the solution of the differential equation with given boundary conditions at $\lambda=0$ is unique. For $\lambda < 0$, for which there are complex eigenvalues of the Hamiltonian, we find a non-perturbative ambiguity. This ambiguity is due to the contribution to the partition sum of states whose energies diverge like $1/|\lambda|$ in the limit $\lambda\to 0$. It is related to the fact that the series in $\lambda$ that defines the partition sum is asymptotic.

The plan of the paper is the following. In section \ref{sec:defspectrum} we determine to all orders in the coupling $\lambda$ the torus partition sum of any theory that satisfies the modular invariance condition (\ref{eq:modularZdef}) and the assumption on the spectrum mentioned above. In particular, we show that this assumption leads to a recursion relation, (\ref{eq:recursionZ}), for the coefficients in the perturbative expansion (\ref{eq:expzz}). We prove that the only solution is the partition sum of a $T\bar T$ deformed CFT.

In section \ref{sec:flow-eq} we use a differential equation that follows from the above recursion relation, (\ref{eq:flow-eq-1}), to study the partition sum non-perturbatively in $\lambda$. We find that for positive $\lambda$, there are no non-trivial non-perturbative effects, while for negative $\lambda$ there is a non-perturbative ambiguity associated with states whose energies go to infinity as $\lambda\to 0$.

For $\lambda > 0$ the partition sum has a Hagedorn singularity. We discuss this behavior and some of its implications in section \ref{sec:hagedorn}.

In section \ref{sec:holog} we discuss the relation of our results to holographic constructions of $T\bar T$ deformed CFTs and of related deformations of large $c$ conformal field theories.

We end in section \ref{sec:conc} with a summary and a discussion of some future directions.

\section{Spectrum from modular invariance}
\label{sec:defspectrum}
In this section, we show that modular invariance (\ref{eq:modularZdef}), and  the qualitative assumption about the spectrum described in the previous section, allows one to uniquely fix the partition sum to all orders in $\lambda$.

The torus partition sum of the undeformed CFT, as a function of $\tau = \tau_1 + i \tau_2$, is given by the standard expression
\begin{align}
\label{eq:zzero}
Z_0(\tau,\bar{\tau})= \tr \left[e^{2\pi i \tau(L_0-{c\over 24})} e^{-2\pi i \btau(\bar{L}_0-{c\over 24})}\right] =\sum_n e^{2\pi i\tau_1 RP_n-2\pi\tau_2 RE_n},
\end{align}
where the sum over $n$ runs over all the eigenstates $|n\rangle$ of the Hamiltonian $H$ and of the spatial momentum $P$ on a circle of radius $R$, and $P_n$ and $E_n$ are the momentum and energy of the state $|n\rangle$, related to the eigenvalues of $L_0$, $\bar L_0$ via
\begin{align}
\label{eq:epn}
(L_0-\overline{L}_0)|n\rangle=RP_n|n\rangle,\qquad \left(L_0+\overline{L}_0-\frac{c}{12}\right)|n\rangle=RE_n|n\rangle.
\end{align}
For any consistent CFT, the partition sum (\ref{eq:zzero}) is modular invariant
\begin{align}
Z_0\left(\frac{a\tau+b}{c\tau+d},\frac{a\bar{\tau}+b}{c\bar{\tau}+d}\right)=Z_0(\tau,\bar{\tau}),
\end{align}
for any integers $a,b,c,d$ with $ad-bc=1$.

We now consider a deformation of the CFT that satisfies the property mentioned in section \ref{sec:1}: the states $|n\rangle$ of the original theory with energies $E_n$ and momenta $P_n$, are deformed at finite $\lambda$ to states $|n\rangle_\lambda$ with energies $\mathcal{E}$ and the same (quantized) momenta,
\begin{align}
\label{eq:depeenn}
E_n\mapsto\mathcal{E}(E_n,P_n,\lambda),\qquad P_n\mapsto P_n,
\end{align}
such that  $\mathcal{E}$ depends on the energy and momentum of the undeformed state $|n\rangle$, and on $\lambda$ via the universal function $\mathcal{E}(E_n,P_n,\lambda)$, (2.4). For now we restrict our attention to states whose energies have a regular Taylor expansion in $\lambda$,
\begin{align}
\label{eq:pertvarEn}
\mathcal{E}(E_n,P_n,\lambda)=\sum_{k=0}^\infty \rE_n^{(k)}\lambda^k=\rE_n^{(0)}+\lambda\,\rE_n^{(1)}+\lambda^2\,\rE_n^{(2)}+\cdots,
\end{align}
where $\rE_n^{(0)}=E_n$ is the undeformed energy (\ref{eq:epn}), and $\rE_n^{(k>0)}$ are functions of $E_n, P_n$ to be determined.

Plugging (\ref{eq:pertvarEn}) into the partition sum
\begin{align}
\label{eq:genpart}
\mathcal{Z}(\tau,\bar{\tau}|\lambda)
=\sum_{n} e^{2\pi i\tau_1 RP_n-2\pi \tau_2 R\mathcal{E}(E_n,P_n,\lambda)}
\end{align}
leads to a Taylor expansion
\begin{align}
\label{eq:expzz}
\mathcal{Z}(\tau,\bar{\tau}|\lambda)=\sum_{k=0}^\infty Z_k\,\lambda^k=Z_0+Z_1\,\lambda+Z_2\,\lambda^2+\cdots
\end{align}
for the perturbed partition sum. Here $Z_0(\tau,\bar{\tau})$ is the undeformed CFT partition sum (\ref{eq:zzero}).

If the deformed CFT contains a single scale, associated with a dimensionful coupling $t$, we can form a dimensionless combination, $\lambda$, from $t$ and an appropriate power of $R$, such that the torus partition sum depends only on the modular parameter $\tau$ and on $\lambda$. Since modular transformations act\footnote{This can be seen, for example, by noting that the area of the torus, $R^2\tau_2$, must be invariant under modular transformations.} on $R$ but do not change $t$, $\lambda$ transforms non-trivially.

We begin with the special case in which $t$ has scaling dimension $(-1,-1)$, leading to the modular transformation (\ref{eq:modularZdef}). This implies that  the  coefficients $Z_p(\tau,\bar\tau)$ in (\ref{eq:expzz}) transform as modular forms of weight $(p,p)$,
\begin{align}
\label{eq:modpp}
Z_p\left(\frac{a\tau+b}{c\tau+d},\frac{a\bar{\tau}+b}{c\bar{\tau}+d}\right)=(c\tau+d)^p(c\bar{\tau}+d)^p\,Z_p(\tau,\bar{\tau}).
\end{align}
We will next show that (\ref{eq:depeenn}), (\ref{eq:pertvarEn}), (\ref{eq:modpp})  determine $Z_p(\tau,\bar\tau)$ uniquely.

To do that one can proceed as follows. Plugging the perturbative expansion of the energies (\ref{eq:pertvarEn}) into the partition sum (\ref{eq:genpart}), one gets explicit expressions for the coefficient functions $Z_p(\tau,\bar\tau)$ in terms of the energy shifts $\rE_n^{(k)}$. The first few of those are:
\begin{align}
\label{eq:Z123}
&Z_1=\sum_n\left(-2\pi R\tau_2\rE^{(1)}_n\right)e^{2\pi i\tau_1 RP_n-2\pi\tau_2 RE_n},\\\nonumber
&Z_2=\sum_n\left(\frac{\tau_2^2}{2}(2\pi R\rE_n^{(1)})^2-2\pi R\tau_2\rE_n^{(2)}\right)e^{2\pi i\tau_1 RP_n-2\pi \tau_2 RE_n},\\\nonumber
&Z_3=\sum_n\left(-\frac{\tau_2^3}{6}(2\pi R\rE_n^{(1)})^3+(2\pi R\tau_2)^2\rE_n^{(1)}\rE_n^{(2)}-2\pi R\tau_2\rE_n^{(3)}\right)e^{2\pi i\tau_1 RP_n-2\pi\tau_2 RE_n}.
\end{align}
Continuing to higher values of $p$, it is easy to see that the expression for $Z_p$ for general $p$ has the following properties:
 \begin{enumerate}
 \item Since $\rE^{(k)}_n$ are functions of the unperturbed energies and momenta, $E_n$ and $P_n$, in expressions such as (\ref{eq:Z123}) they can be replaced by differential operators in $\tau$ and $\bar\tau$, using
 \begin{align}
2\pi RE_n\mapsto-\partial_{\tau_2}=\frac{1}{i}(\partial_{\tau}-\partial_{\bar{\tau}}),\qquad
2\pi i RP_n\mapsto\partial_{\tau_1}=\partial_{\tau}+\partial_{\bar{\tau}}.
\end{align}
\item After doing that, $Z_p$ takes the general form
\begin{align}
\label{eq:Zpexpand}
Z_{p}=\left[\tau_2^p\widehat{\mathcal{O}}_1^{(p)}(\partial_{\tau},\partial_{\bar{\tau}})
+\tau_2^{p-1}\widehat{\mathcal{O}}_2^{(p)}(\partial_{\tau},\partial_{\bar{\tau}})+\cdots
+\tau_2\widehat{\mathcal{O}}_p^{(p)}(\partial_{\tau},\partial_{\bar{\tau}})\right]Z_0(\tau,\bar\tau),
\end{align}
where $\widehat{\mathcal{O}}_j^{(p)}(\partial_{\tau},\partial_{\bar{\tau}})$ are differential operators, that encode the information about the energy shifts  $\rE^{(k)}_n$.
\item $\widehat{\mathcal{O}}_1^{(p)}$ is fixed by  $\rE^{(1)}_n$ (and vice versa). $\widehat{\mathcal{O}}_2^{(p)}$ is fixed by  $\rE^{(1)}_n$ and $\rE^{(2)}_n$. In general, $\widehat{\mathcal{O}}_k^{(p)}$ is fixed by $\rE^{(j)}_n$, with $j=1,2,\cdots, k$.
\end{enumerate}

We can use the properties listed above to show that given $Z_0$, there is a unique $Z_p$ that satisfies all the constraints. We will do this using induction, by showing that if $Z_0,\cdots,Z_p$ have been uniquely determined,  $Z_{p+1}$ can be determined as well.

Before discussing the general case, it is useful to consider the case $p=1$ in (\ref{eq:Zpexpand}). We are looking for an operator
$\widehat{\mathcal{O}}_1^{(1)}$, such that
\begin{align}
\label{eq:Zone}
Z_1=\tau_2\widehat{\mathcal{O}}_1^{(1)}(\partial_{\tau},\partial_{\bar{\tau}})Z_0(\tau,\bar\tau)
\end{align}
is (using \eqref{eq:modpp}) a modular form of weight $(1,1)$ for any modular invariant $Z_0(\tau,\bar\tau)$. To find $\widehat{\mathcal{O}}_1^{(1)}$, it is useful to recall the modular covariant derivatives \cite{Zagier}
\begin{align}
\label{eq:defdd}
\textsf{D}_{\tau}^{(k)}=\partial_{\tau}-\frac{ik}{2\tau_2},\qquad \textsf{D}_{\bar{\tau}}^{(k)}=\partial_{\bar{\tau}}+\frac{ik}{2\tau_2}.
\end{align}
Acting with $\textsf{D}_{\tau}^{(k)}$ on a modular form of weight $(k,k')$ gives a modular form of weight $(k+2, k')$. Similarly, $\textsf{D}_{\bar{\tau}}^{(k')}$ increases the weight of such a modular form to $(k, k'+2)$. It is also useful to recall that $\tau_2$ is a modular form of weight $(-1,-1)$.

Looking back at (\ref{eq:Zone}), we see that we need to find a differential operator $\widehat{\mathcal{O}}_1^{(1)}$, such that $\widehat{\mathcal{O}}_1^{(1)}(\partial_{\tau},\partial_{\bar{\tau}})Z_0(\tau,\bar\tau)$ is a modular form of weight $(2,2)$. Clearly, the unique operator with these properties is
\begin{align}
\label{eq:defalpha}
\widehat{\mathcal{O}}_1^{(1)}(\partial_{\tau},\partial_{\bar{\tau}})=\alpha\partial_{\tau}\partial_{\bar{\tau}},
\end{align}
where $\alpha$ is an arbitrary constant. Indeed, according to (\ref{eq:defdd}) with $k=0$, $\widehat{\mathcal{O}}_1^{(1)}(\partial_{\tau},\partial_{\bar{\tau}})Z_0(\tau,\bar\tau)$ is in this case a modular form of weight $(2,2)$, while acting with additional derivatives with respect to $\tau$ and/or $\bar\tau$ gives rise to non-zero contributions that transform as $(k,k')$ forms with $k$ and/or $k'$ larger than two.

The constant $\alpha$ does not play a role in the discussion, as it can be absorbed into the definition of the coupling $\lambda$ (see (\ref{eq:expzz})). We will set it to one below.

We are now ready to discuss the general induction step, going from $p$ to $p+1$ (with $p>0)$. As explained earlier, assuming that $Z_0, Z_1,\cdots, Z_p$ have been uniquely fixed means that the energy shifts $\rE^{(j)}_n$, with $j=1,2,\cdots, p$, have been fixed as well. In the expression (\ref{eq:Zpexpand}) for $Z_{p+1}$, the operators $\widehat{\mathcal{O}}_k^{(p+1)}$ with $k=1,2,\cdots, p$ are thus uniquely determined, and the only unknown operator is $\widehat{\mathcal{O}}_{p+1}^{(p+1)}$.

Suppose there are two such operators that satisfy all the constraints. Since each of them gives rise to a $Z_{p+1}$ with the right modular transformation properties, (\ref{eq:modpp}), the difference between them should also transform as a modular form of weight $(p+1,p+1)$. However, in the difference, all the terms that go like powers of $\tau_2$ larger than one in (\ref{eq:Zpexpand}) cancel, and we conclude that there must exist an operator $\delta\widehat{\mathcal{O}}_{p+1}^{(p+1)}(\partial_{\tau},\partial_{\bar{\tau}})$, such that $\delta\widehat{\mathcal{O}}_{p+1}^{(p+1)}(\partial_{\tau},\partial_{\bar{\tau}})Z_0(\tau,\bar\tau)$ is a modular form of weight $(p+2,p+2)$.

To see that such an operator does not exist, consider the action of $\partial_\tau$, $\partial_{\bar\tau}$ on modular forms $f_{k,\bar{k}}(\tau,\bar{\tau})$ of general weight $(k,\bar k)$ (see (\ref{eq:defdd})),
\begin{align}\label{eq:cov-D}
&\partial_{\tau}f_{k,\bar{k}}(\tau,\bar{\tau})=\textsf{D}_{\tau}^{(k)}f_{k,\bar{k}}(\tau,\bar{\tau})
+\frac{ik}{2\tau_2}f_{k,\bar{k}}(\tau,\bar{\tau}),\\\nonumber
&\partial_{\bar{\tau}}f_{k,\bar{k}}(\tau,\bar{\tau})=\textsf{D}_{\bar{\tau}}^{(\bar{k})}f_{k,\bar{k}}(\tau,\bar{\tau})
-\frac{i\bar{k}}{2\tau_2}f_{k,\bar{k}}(\tau,\bar{\tau}).
\end{align}
The first line says that acting with  $\partial_{\tau}$ on a modular form of weight $(k,\bar{k})$ gives a linear combination of modular forms of weights $(k+2,\bar{k})$ and $(k+1,\bar{k}+1)$. Similarly, acting with $\partial_{\bar{\tau}}$ gives a linear combination of forms of weights $(k,\bar{k}+2)$ and $(k+1,\bar{k}+1)$. In particular, the total (left $+$ right) weight always increases by two units, but the individual weights of different contributions are in general different.

The fact that $\delta\widehat{\mathcal{O}}_{p+1}(\partial_{\tau},\partial_{\bar{\tau}})Z_0(\tau,\bar\tau)$ must be a modular form of weight $(p+2,p+2)$ implies that the operator $\delta\widehat{\mathcal{O}}_{p+1}(\partial_{\tau},\partial_{\bar{\tau}})$ must contain a combined total of $p+2$ derivatives with respect to $\tau$ and $\bar\tau$. It is easy to see that an arbitrary linear combination of all such terms has, in addition to the desired $(p+2,p+2)$ form, other contributions from $(k,\bar{k})$ forms with $k\not=\bar k$ and $k+\bar k=2p+4$, which do not vanish for general $Z_0$. Thus, $\delta\widehat{\mathcal{O}}_{p+1}$ must vanish, and we conclude that $Z_{p+1}$ is also unique.

Having established that the form of the $Z_p$'s is unique given $Z_0$ (up to the freedom of rescaling $\lambda$ in (\ref{eq:expzz})), it is natural to ask whether they can be computed in closed form. It turns out that a useful ansatz is the recursion relation
\begin{align}
\label{eq:recursion}
Z_{p+1}=d_p\left(\tau_2\textsf{D}_{\tau}^{(p)}\textsf{D}_{\bar{\tau}}^{(p)}-\frac{b_p}{\tau_2}\right)Z_p,
\end{align}
where $d_p$, $b_p$ are constants to be determined. If $Z_p$ is a weight $(p,p)$ modular form, then $Z_{p+1}$
(\ref{eq:recursion}) is (by construction) a $(p+1,p+1)$ form, as expected (\ref{eq:modpp}). The constants  $d_p$, $b_p$ can be determined as follows:
\begin{itemize}
\item $d_p$ can be determined by comparing the coefficients of $\tau_2^{p+1}$ on the left and right hand sides of (\ref{eq:recursion}), using (\ref{eq:Zpexpand}) and the fact that for each energy level $\widehat{\mathcal O}_1^{(p)}$ gives a factor of $(-2 \pi R \rE_n^{(1)})^p / p!$;  one finds $d_p=1/(p+1)$.
\item $b_p$ can be determined by demanding that when $Z_p$ does not have a term that goes like $\tau_2^0$ in the expansion  (\ref{eq:Zpexpand}), neither should $Z_{p+1}$. A short calculation leads to $b_p=p(p+1)/4$.
\end{itemize}

Thus, we conclude that the coefficients $Z_p$ must satisfy the recursion relation
\begin{align}
\label{eq:recursionZ}
Z_{p+1}=\frac{\tau_2}{p+1}\left(\textsf{D}_{\tau}^{(p)}\textsf{D}_{\bar{\tau}}^{(p)}-\frac{p(p+1)}{4\tau_2^2}\right)Z_p.
\end{align}
In particular, all $Z_{p>0}$ are uniquely determined by the unperturbed partition sum $Z_0$ (\ref{eq:zzero}).

As discussed above, one can use  (\ref{eq:recursionZ}) to determine the energy shifts $\rE^{(j)}_n$  \eqref{eq:pertvarEn}.
The first few of these are
\begin{align}
\rE^{(1)}_n &=  -\frac{\pi R}{2} (E_n^2 -P_n^2) ,\\
\rE^{(2)}_n &=  \frac{\pi^2R^2}{2} (E_n^2 -P_n^2)  E_n\nonumber ,\\
\rE^{(3)}_n &=  -\frac{\pi^3R^3}{8} (E_n^2 -P_n^2)  (5E_n^2 -P_n^2). \nonumber
\end{align}
These are the first terms in the expansion of the energy spectrum of $T\bar T$ deformed CFTs,
\begin{align} \label{eq:deformed-spectrum}
\mathcal{E}_n\equiv\mathcal{E}(E_n,P_n,\lambda)=\frac{1}{\pi \lambda R}\left(\sqrt{1+2\pi\lambda RE_n+\lambda^2\pi^2 R^2 P_n^2}-1\right),
\end{align}
where we used the conventions of \cite{Cavaglia:2016oda},\footnote{Our convention for the deformed energy is slightly different from the one in \cite{Cavaglia:2016oda}, $2\pi\mathcal{E}_{\text{here}}=\mathcal{E}_{\text{there}}$.} with $\lambda=4 t/R^2$. Note that for $\lambda$ positive and sufficiently small, these energies are real, while for any negative $\lambda$ the spectrum arising from large enough energies $E_n$ becomes complex, so the theory cannot be unitary (the Hamiltonian is not Hermitian).

Another way to see that the recursion relation (\ref{eq:recursionZ}) gives rise to the spectrum of a $\ttb$ deformed CFT is to note that it is identical to the one found in  \cite{Datta:2018thy}, from the diffusion equation for the partition function of that model \cite{Cardy:2018sdv}. We will also see in the next section that this recursion relation gives rise to the inviscid Burgers equation for the spectrum of a $T\bar T$ deformed CFT found in \cite{Smirnov:2016lqw,Cavaglia:2016oda}.

So far in this section we assumed that the partition sum (\ref{eq:genpart}) has an expansion in integer powers of the dimensionless coupling $\lambda$, as in (\ref{eq:expzz}), where $\lambda$ transforms as \eqref{eq:modularZdef}. The motivation for this was that $\lambda$ is proportional to the coupling $t$, which we took to have dimension $(-1,-1)$. A natural question is whether there is another class of theories that satisfies our requirements, in which the coupling has a different dimension, such that the dimensionless coupling $\lambda$ has a different modular weight.

If such a class existed, it could be studied in our formalism by defining a coupling $\hat\lambda$ that has the same weight as in our analysis, and writing the physical dimensionless coupling of the theory as $\lambda=\hat\lambda^a$ with some real number, $a$. Thus, in terms of our analysis, the question becomes whether there is another class of partition sums satisfying our requirements, in which the leading correction to the CFT partition sum $Z_0$ is $Z_1\hat\lambda^a$, with $a\not=1$.

Repeating the analysis from before, we can write $Z_1$ in the form  (\ref{eq:Zone}), and  since $Z_1$ must be a modular form of weight $(a,a)$, $\widehat{\mathcal{O}}_1(\partial_{\tau},\partial_{\bar{\tau}})Z_0(\tau,\bar\tau)$ must have weight $(a+1,a+1)$. For positive integer $a$, we have seen before that this is impossible. It is easy to see that it is impossible for non-integer $a$ as well, due to the fact that the operator $\widehat{\mathcal{O}}_1(\partial_{\tau},\partial_{\bar{\tau}})$ must have, by construction, a good Taylor expansion in its arguments. Negative values of $a$ (corresponding to relevant perturbations of a CFT) can be ruled out in a similar way.

Thus, we conclude that perturbatively in any (single) dimensionful coupling, a $T\bar T$ deformed CFT is the only solution to the requirements we imposed.

 \section{Non-perturbative analysis}
 \label{sec:flow-eq}

In section \ref{sec:defspectrum} we determined the partition sum \eqref{eq:genpart} to all orders in the coupling $\lambda$. It is natural to ask what happens beyond perturbation theory. The first question we need to address is what  we mean by non-perturbative contributions to the partition sum from the general perspective of the previous sections.

We saw that the coefficient functions $Z_p$ in the expansion \eqref{eq:expzz} satisfy the recursion relation \eqref{eq:recursionZ}. This recursion relation can be summarized in a compact way as a differential equation for the partition sum $\mathcal{Z}(\tau,\bar{\tau}|\lambda)$ \eqref{eq:expzz},
\begin{align}\label{eq:flow-eq-1}
 	\pd_\lambda \mathcal{Z}(\tau,\bar{\tau}|\lambda)= \left[ \tau_2 \pd_\tau\pd_{\btau} + \frac{1}{2}\left(i\,  (\pd_\tau-\pd_{\btau})- \frac{1}{\tau_2}\right) \lambda \pd_\lambda \right] \mathcal{Z}(\tau,\bar{\tau}|\lambda).
 \end{align}
Indeed, plugging the expansion  \eqref{eq:expzz} into \eqref{eq:flow-eq-1} gives the recursion relation \eqref{eq:recursionZ}. Alternatively, plugging \eqref{eq:genpart} into \eqref{eq:flow-eq-1}, and comparing the coefficients of particular terms in the sum over $n$ on the left and right hand sides, gives an ODE in $\lambda$. This ODE is equivalent to the inviscid Burgers equation for the deformed energies  derived in \cite{Smirnov:2016lqw,Cavaglia:2016oda}, which is indeed solved by \eqref{eq:deformed-spectrum}.

A natural non-perturbative completion of the construction of the previous sections is to take the partition sum  $\mathcal{Z}(\tau,\bar{\tau}|\lambda)$ to obey the differential equation \eqref{eq:flow-eq-1} with the boundary condition
\begin{align}\label{eq:boundcond}
\mathcal{Z}(\tau,\bar{\tau}|0)=Z_0(\tau,\bar\tau), 	
 \end{align}
where $Z_0(\tau,\bar\tau)$ is the partition sum of the original CFT \eqref{eq:zzero}.

In the context of a $T\bar T$ deformed CFT,  an identical equation (written in a different form) was derived from the path integral in  \cite{Cardy:2018sdv}, so that this deformation gives an example of such a non-perturbative completion\footnote{In fact, \cite{Cardy:2018sdv} showed that the same equation holds also for general $T{\bar T}$ deformed theories (not necessarily conformal), such that these theories also obey \eqref{eq:modularZdef}, even though they have more than one mass scale.}. From our more general perspective, which does not assume a priori that we are dealing with a $T\bar T$ deformed CFT, it is a natural non-perturbative completion.

We now ask what non-perturbative effects do \eqref{eq:flow-eq-1}, \eqref{eq:boundcond} describe. A useful way of thinking about this is the following. As mentioned above, the differential equation \eqref{eq:flow-eq-1} determines the spectrum of energies  \eqref{eq:deformed-spectrum}. In that equation we took the positive branch of the square root, because we wanted the energies to satisfy the boundary condition $\mathcal{E}(E_n,P_n,0)=E_n$.

Denoting the energies  by $\mathcal{E}_n$ in \eqref{eq:deformed-spectrum} by $\mathcal{E}_n^{(+)}$, the negative branch of the square root gives another set of energies, $\mathcal{E}_n^{(-)}$, related to those in \eqref{eq:deformed-spectrum} via the relation
\begin{align}\label{eq:epsilonpm}
\mathcal{E}_n^{(+)}+ \mathcal{E}_n^{(-)}=-\frac{2}{\pi\lambda R}.	
 \end{align}
In particular, while $\mathcal{E}_n^{(+)}$ have a finite limit as $\lambda\to 0$, $\mathcal{E}_n^{(-)}$ behave as $\lambda \to 0$ like
\begin{align}\label{eq:epsilonminus}
 \mathcal{E}_n^{(-)}\sim-\frac{2}{\pi\lambda R}~.
 \end{align}
For positive (negative) $\lambda$ they go to minus (plus) infinity. This will lead to a difference in the analysis of the partition function between positive and negative values of $\lambda$, even though no such difference appeared in the perturbative expansion of the previous section.

The key point for our purposes is that the differential equation \eqref{eq:flow-eq-1} is linear in $\mathcal{Z}$. Thus, it holds separately for the contribution of any specific state to $\mathcal{Z}$, and is valid for both branches of the spectrum. The perturbative contribution to the partition sum studied in section \ref{sec:defspectrum} is obtained by plugging into \eqref{eq:genpart} the energies $\mathcal{E}_n^{(+)}$. If, on the other hand, we take for the spectrum some other states with energies $\tilde{\mathcal{E}}_n=\tilde{\mathcal{E}}_n^{(-)}$ in \eqref{eq:genpart}, we get a partition sum that goes like
\begin{align}
\label{eq:genpartnonpert}
\mathcal{Z}_{\rm np}(\tau,\bar{\tau}|\lambda)
=e^{4\tau_2/\lambda}\sum_{n} e^{2\pi i\tau_1 RP_n+2\pi \tau_2 R\tilde{\mathcal{E}}_n^{(+)}},
\end{align}
where we used \eqref{eq:epsilonpm}. The $\tilde{\mathcal{E}}_n^{(+)}$ here are given by \eqref{eq:deformed-spectrum} with some energies $\tilde{E}_n$.

Note that:
\begin{enumerate}
\item The coefficient of the exponential in \eqref{eq:genpartnonpert} has a good Taylor expansion in $\lambda$. It can be thought of as obtained from some spectrum \eqref{eq:deformed-spectrum} by taking $\lambda\mapsto -\lambda$, ${\tilde E}\mapsto -{\tilde E}$, ${\tilde {\mathcal{E}}}\mapsto -{\tilde {\mathcal{E}}}$, which preserves the form of our equations. Note that it is natural to take most of the $\tilde{\mathcal{E}}_n^{(+)}$'s in \eqref{eq:genpartnonpert} to be negative, so that the sum over $n$ converges.
\item The exponential in \eqref{eq:genpartnonpert} is modular invariant by itself, so that $\mathcal{Z}_{\rm np}$ is modular invariant if and only if the spectrum ${\tilde E}_n$ corresponds to a modular invariant conformal field theory.
\item The exponential in \eqref{eq:genpartnonpert} diverges badly as $\lambda\to 0^+$. Hence, for positive $\lambda$, such a term is forbidden by the boundary condition \eqref{eq:boundcond}. This is related to the fact that the energies
$\mathcal{E}_n^{(-)}$ go to $(-\infty)$ as $\lambda\to 0$ in this case.
\item For negative $\lambda$, \eqref{eq:genpartnonpert}  goes rapidly to zero as $\lambda\to 0^{-}$. It corresponds to a non-perturbative contribution to the partition sum $\mathcal{Z}(\tau,\bar{\tau}|\lambda)$ that solves \eqref{eq:flow-eq-1}, with extra states whose energies go to $+\infty$ as $\lambda \to 0^-$.
\item
The non-perturbative contribution \eqref{eq:genpartnonpert} to $\mathcal{Z}$ corresponds to an arbitrary modular invariant partition sum multiplying the exponential. In particular, it need not have anything to do with $Z_0$ \eqref{eq:boundcond}. We have a one-to-one correspondence between non-perturbative solutions to \eqref{eq:flow-eq-1} (for a given CFT at $\lambda=0$) and independent modular-invariant CFTs.
\end{enumerate}

The above discussion can be concisely summarized by considering the following ansatz for the non-perturbative solution of
\eqref{eq:flow-eq-1}:
\begin{align}\label{eq:otherSol}
 	\mathcal{Z}_{\rm np} = X(\tau, \btau|\lambda)\, \exp\left[4\tau_2 \over \lambda \right].
 \end{align}
 Plugging this into \eqref{eq:flow-eq-1}, we find that the prefactor $X$ satisfies
 \begin{align}\label{eq:flow-eq-p}
 	\pd_\lambda X  = - \left[ \tau_2 \pd_\tau\pd_{\btau} + \frac{1}{2}\left(i\,  (\pd_\tau-\pd_{\btau})- \frac{1}{\tau_2}\right) \lambda \pd_\lambda \right] X.
 \end{align}
 As anticipated by the discussion above, this is the same as the original equation \eqref{eq:flow-eq-1}, with $\lambda \mapsto -\lambda$. Comparing \eqref{eq:genpartnonpert} and \eqref{eq:flow-eq-p}, we see that $X$ in the latter is the pre-exponential factor in the former. In particular, it has a smooth limit as $\lambda\to 0$.

Thus, we conclude that for $\lambda>0$, the solution of \eqref{eq:flow-eq-1}, \eqref{eq:boundcond} does not have any non-perturbative ambiguities, while for $\lambda<0$ it has an ambiguity of the form \eqref{eq:otherSol}. This ambiguity is parametrized by a choice of a modular invariant function $X_0=X(\tau,\btau|0)$, which provides the boundary condition for
\eqref{eq:flow-eq-p}.

The form of the non-perturbative contribution to the partition sum suggests that the perturbative series \eqref{eq:expzz} is asymptotic, with the appropriate large order growth. More precisely, one expects that  for large $p$ we have
\begin{align}
\label{highorder}
 	Z_p(\tau,\btau) \ \sim \ \frac{(p-1)!}{(-4\tau_2)^p}\,  Y (\tau,\btau),
 \end{align}
 where $Y (\tau,\btau)$ is a modular invariant function.
 From the perspective of \eqref{eq:recursionZ} this is the statement that the solution of the recursion relation approaches at large $p$ a zero mode of the operators $\textsf{D}_{\tau}^{(p)}$, $\textsf{D}_{\bar{\tau}}^{(p)}$, which is not unreasonable. In particular, note that for $Y=1$, \eqref{highorder} gives a solution to this recursion relation.

 \section{Some properties of the torus partition function}
 \label{sec:hagedorn}
 In this section we briefly comment on some properties of the deformed partition function $\mathcal{Z}(\tau,\bar{\tau}|\lambda)$.

Let us start with the case $\lambda > 0$, where the deformed energies are real for small enough $\lambda$. As is shown in \cite{Giveon:2017nie,Datta:2018thy}, for $\lambda > 0$ and a CFT of central charge $c$, the density of states of the deformed theory interpolates between Cardy behavior, $\rho(E)\sim e^{\sqrt{ {2\pi R c \over 3} E}}$ in an intermediate range of energies (which is only present for $\lambda \ll 1$), and Hagedorn behavior, $\rho(E)\sim e^{ 2\pi R E  \sqrt{\pi c\lambda  \over  6}}$ at asymptotically large energies.

Consider, for simplicity, the partition sum (\ref{eq:genpart}), for $\tau_1$=0. The Hagedorn behavior of the asymptotic spectrum implies that the partition sum is convergent only for
\begin{align}\label{eq:ttone}
 	\tau_2>\tau_2^{\text{H}}(\lambda),
 \end{align}
with
\begin{align}\label{eq:tttwo}
 	\tau_2^{\text{H}}(\lambda)=\sqrt{\frac{\pi c\lambda}{6}}.
 \end{align}
Modular invariance, (\ref{eq:modularZdef}), implies that
\begin{align}\label{eq:ttthree}
 	\mathcal{Z}(\tilde\tau_2,\tilde\lambda)=Z(\tau_2,\lambda),\;\;{\rm with}\;\;\tilde\tau_2=\frac{1}{\tau_2},\;\;\tilde\lambda=\frac{\lambda}{\tau_2^2}.
 \end{align}
The convergence requirement (\ref{eq:ttone}) is mapped by (\ref{eq:ttthree}) to the condition
\begin{align}\label{eq:ttfour}
\frac{\pi c\tilde\lambda}{6}<1.
 \end{align}
Thus, the partition sum on a rectangular torus is only well-defined when both sides of the torus are larger than $2\pi \sqrt{\frac{2\pi c t}{3}}$.

It is useful to note that:
\begin{enumerate}
\item The Hagedorn singularity, that in terms of the original variables $(\tau_2,\lambda)$, happens at a particular value of $\tau_2$ that depends on $\lambda$, (\ref{eq:ttone}), (\ref{eq:tttwo}), happens in the dual variables at a particular value of the dual coupling $\tilde\lambda$, (\ref{eq:ttfour}), for all values of the dual modulus $\tilde\tau_2$.
\item There is an independent reason to require the condition (\ref{eq:ttfour}) on the coupling. Looking back at (\ref{eq:deformed-spectrum}), we see that this condition is necessary for the $SL(2,\mathbb{R})$ invariant vacuum of the original CFT, which has $E_0R=-c/12$ and $P_0=0$ (\ref{eq:epn}), to have a real energy $\mathcal{E}_0$ in the deformed theory. For larger values of $\lambda$, or equivalently smaller values of $R$ for a given $t$, this energy becomes complex. The condition \eqref{eq:ttfour} can be thought of as the requirement that the coupling at the scale $R$ be sufficiently weak.
\item The above discussion is reminiscent of the usual relation between the high energy density of states and the mass of the lowest lying state winding around Euclidean time in a free string theory at finite temperature.
\item The modular parameter of the torus, $\tau$, can be restricted to a single fundamental domain, e.g. the standard domain $|\tau|\ge 1$, $\tau_1\in [-1/2,1/2]$. If we impose the condition (\ref{eq:ttfour}) on the coupling and use the fact that in that domain, for $\tau_1=0$ one has $\tau_2\geq 1$, we see that the Hagedorn singularity \eqref{eq:tttwo} is never reached.
\end{enumerate}

For $\lambda < 0$, the spectrum of energies \eqref{eq:deformed-spectrum} is complex for large enough undeformed energies $E_n$. This leads to the torus partition sum also being complex (even for $\tau_1=0$). The interpretation of such non-unitary deformations of unitary theories is not clear. However, it is interesting to note that there is one specific non-perturbative completion for which the partition function becomes real. It corresponds to accompanying each state with energy $\mathcal{E}_n^{(+)}$ by a state of energy $\mathcal{E}_n^{(-)}$, or in other words to choosing the ${\tilde E}_n$ appearing in \eqref{eq:genpartnonpert} to be ${\tilde E}_n = -E_n$. This seems to be a natural UV completion of the partition function, but from our point of view it is not clear why it is preferred compared to others.

\section{Relation to holography}
\label{sec:holog}

Many two dimensional CFT's are related via holographic duality to vacua of string theory on $AdS_3$. After the original papers \cite{Smirnov:2016lqw,Cavaglia:2016oda}, there has been some work on the fate of these dual pairs after a $T\bar T$ deformation \cite{McGough:2016lol,Cottrell:2018skz,Kraus:2018xrn,Asrat:2017tzd,Giveon:2017myj,Giveon:2017nie,Giribet:2017imm,Babaro:2018cmq,Chakraborty:2018kpr}. In this section we comment on the interpretation of our results in that context, leaving a more detailed discussion to future work. There are two distinct holographic constructions that we discuss in turn below.

The first involves starting with an $AdS_3$ vacuum of string theory, which is weakly coupled corresponding to a large $c$ CFT, and turning on the $t T {\bar T}$ deformation in the dual CFT. Such a deformation which is quadratic in CFT operators is known as a ``double-trace'' deformation. At leading order in $t$ and in $1/c$ the general rules of the AdS/CFT correspondence \cite{Witten:2001ua,Berkooz:2002ug} imply that it modifies the boundary condition for the graviton at the boundary of $AdS_3$. More precisely, this is true when one takes the large $c$ limit keeping fixed the combination $(t \cdot c)$ (note that this is the same combination appearing in the Hagedorn temperature; see (\ref{eq:tttwo})). Because $t \sim 1/c$, the changes in the spectrum of light states in the bulk are very small, but the changes in the energies of black hole states with $E \sim c$ can be large. The description as a deformation of the boundary condition reproduces correctly the perturbation expansion of correlation functions and other observables in $(t\cdot c)$, but it is not clear how to generalize it to finite values of $(t\cdot c)$, or to higher orders in the expansion in powers of $t$ or $1/c$. Since it is an irrelevant deformation, this requires significant changes in the behavior near the boundary of $AdS_3$.

The leading order description above holds for both signs of $t$. For $t>0$ where the spectrum is unitary, there are no known candidates for the dual holographic description at finite $t$; our considerations suggest that this dual should be unique. For $t<0$, where the spectrum is non-unitary, a suggestion for the holographic dual at finite $t$ appeared in \cite{McGough:2016lol} and was analyzed further in \cite{Kraus:2018xrn,Cottrell:2018skz,Taylor:2018xcy,Donnelly:2018bef,Chen:2018eqk}. This suggestion involves putting a finite cutoff on the radial direction of $AdS_3$, at a position related to $t$. This correctly reproduces many features of the $T{\bar T}$ deformation, but when there are more fields (beyond the graviton) in the bulk, such a finite cutoff corresponds to a much more complicated deformation involving many different double-trace operators \cite{Kraus:2018xrn}. It is not clear if such a deformation, which involves many dimensionful coupling constants, has any special features, and it is not directly related to our discussion here. In any case, our analysis implies that there can be many different UV completions of the $T {\bar T}$ deformation for this sign of $t$, and it is not clear which, if any, could be related to a finite cutoff in the bulk. If we accept the relation to a finite cutoff, it is tempting to suggest that perhaps this freedom corresponds to different choices of the fields living beyond the cutoff, while not modifying the physics inside the cutoff.

In addition, the authors of \cite{McGough:2016lol} suggested that the spectrum of energies of the deformed theory should be cut off at the value of the energy where the energy spectrum \eqref{eq:deformed-spectrum} becomes complex, and that all higher energy states should be removed. Such a truncation is not modular invariant by itself. However, note that already for $AdS_3$, modular transformations exchange different gravitational solutions in the bulk (which are all locally $AdS_3$) \cite{Dijkgraaf:2000fq,Manschot:2007ha}, in which different cycles of the torus shrink to zero in the bulk. This suggests that one could make the finite cutoff prescription modular invariant by adding to its truncated partition function all of its $SL(2,\mathbb{Z})$-transforms. This would correspond to taking into account all these different gravitational solutions (with a finite cutoff) in the bulk. However, there is no reason to believe that these additional contributions would correspond to a consistent spectrum (namely, that they can be written as a sum of the form \eqref{eq:genpart}), so the meaning of this suggestion is not clear.

A second holographic construction, studied in \cite{Asrat:2017tzd,Giveon:2017myj,Giveon:2017nie,Chakraborty:2018kpr,Giribet:2017imm,Babaro:2018cmq}, involves deforming an $AdS_3/CFT_2$ dual pair by adding to the Lagrangian of the CFT a ``single-trace'' operator of dimension $(2,2)$, $D(x,\bar x)$ \cite{Kutasov:1999xu}, which has many features in common with $T\bar T$. From the $AdS_3$ point of view, it corresponds to deforming the geometry from $AdS_3$ to a certain background known as ${\cal M}_3$ \cite{Giveon:1999zm,Giveon:2017nie}. In the worldsheet description of string theory on $AdS_3$ (with NS $B$-field) it corresponds to a null current-current deformation of $AdS_3$ \cite{Forste:1994wp,Giveon:2017nie}.

The geometry of ${\cal M}_3$ depends on the sign of the deformation parameter $t$. For $t>0$, one finds a smooth asymptotically linear dilaton flat three-dimensional space-time,  compactified on a circle with radius $R$, and capped in the infrared region by a locally $AdS_3$ space. We will refer to this background as ${\cal M}_3^{(+)}$.

For $t<0$, the background, which we will denote by ${\cal M}_3^{(-)}$, looks as follows. In the infrared region in the radial coordinate, it approaches $AdS_3$. As one moves towards the UV, the geometry is deformed, and at some value of the radial coordinate, that depends on $t$, one encounters a singularity. The region between the IR $AdS_3$ and the singularity looks like the region between the horizon and the singularity of a black hole. Proceeding past the singularity, the geometry approaches a linear dilaton spacetime. From the point of view of an observer living in that spacetime, the singularity in question is naked. Also, the role of space and time on the boundary are flipped when passing the singularity. Thus, the region past the singularity has closed timelike curves.

While the backgrounds ${\cal M}_3^{(+)}$ and ${\cal M}_3^{(-)}$ look rather different, their constructions in string theory are very similar. As described in \cite{Giveon:2017myj}, the worldsheet theory corresponding to both can be obtained via null gauging of the worldsheet CFT on $\mathbb{R}\times\mathbb{S}^1\times AdS_3$. For $t>0$ $(t<0)$, the gauging involves an axial (vector) symmetry. Therefore, it is natural to expect both of them to give rise to good string backgrounds.

To relate string theory in the deformed backgrounds ${\cal M}_3^{(\pm)}$ to the discussion of this paper, we need to understand the role of the deformation operator $D$ in the CFT dual of string theory on $AdS_3$. In general, the CFT dual to string theory on $AdS_3$ is not well understood, but there is a partial picture that is sufficient for our purposes. We next briefly review this picture and discuss its implications for our case.

The spectrum of string theory on $AdS_3$ includes strings winding around the spatial circle on the boundary and carrying some momentum in the radial direction. Such strings are well described by the symmetric orbifold $M^N/S_N$ \cite{Argurio:2000tb,Giveon:2005mi},
 where $M$ is the theory describing a single string, and $N$ is related to the string coupling, $N \sim 1/g_s^2$. From the point of view of this description, the operator $D$ can be thought of as $\sum_{i=1}^N (T\bar T)_i$, where $(T \bar T)_i$ is the $T{\bar T}$ deformation in the $i$'th copy of $M$. Thus, the single trace deformation studied in \cite{Asrat:2017tzd,Giveon:2017myj,Giveon:2017nie} corresponds from this point of view to the orbifold $M_t^N/S_N$, where $M_t$ is a $T\bar T$ deformed version of the block $M$.

Many aspects of the discussion of this paper have a natural interpretation in the above string theory construction. For example, we found that for $t>0$, the spectrum of the theory does not receive non-perturbative corrections. This is natural in the string theory construction since  ${\cal M}_3^{(+)}$ is a smooth space. An explicit calculation shows that the states in string theory on ${\cal M}_3^{(+)}$ described by the symmetric product do indeed have a smooth limit as $t\to 0^+$.

On the other hand, for $t<0$ we found that the partition sum of the theory has a non-perturbative ambiguity, parametrized by \eqref{eq:otherSol}, \eqref{eq:flow-eq-p}, which corresponds to states with energies that diverge as $t\to 0^-$. It would be interesting to understand these and other features of the field theory discussion from the string theory perspective. It is tempting to speculate that states whose energies have a good perturbative limit correspond in the bulk to wavefunctions that in some sense live in the region between the horizon and the singularity, while those  whose energies diverge in the limit \eqref{eq:epsilonminus} live in the region beyond the singularity. 
Analyzing this could shed light on whether the singularity of the space-time ${\cal M}_3^{(-)}$ is resolved in string theory, and how.
We hope to return to this subject in future work.

\section{Discussion}
\label{sec:conc}

In this paper we studied the torus partition sum of a two dimensional quantum field theory obtained by an irrelevant perturbation of a CFT. We showed that modular invariance, together with the requirement that the energies of states in the perturbed theory depend only on the energies and momenta of the original CFT and on the coupling, places strong constraints on the spectrum. In particular, it fixes it to be that of a $T\bar T$ deformed CFT to all orders in the coupling. In a natural non-perturbative completion, for one sign of the coupling the spectrum is uniquely fixed, while for the other there are non-perturbative ambiguities, that we described.

From the point of view of our paper, all of these non-perturbative (UV) completions of the $t<0$ theory, labeled by an independent modular invariant CFT partition function, are equally valid, and could correspond to a consistent field theory on a torus.  It would be interesting to understand if there are additional constraints that should be imposed on these completions. For instance, these could come from requiring consistency of the theory on higher genus Riemann surfaces, or from the existence and consistency of correlation functions of some local operators (like the energy-momentum tensor itself). Note that in any case these theories are non-unitary, limiting the possible consistency requirements.

The theories we described can be defined in terms of their explicit spectrum of states (with a specific choice made for $t<0$), or, perturbatively in $t$, as $T{\bar T}$ deformations.
Ideally, we would like to have an independent construction of the deformed theory on a torus, which does not rely on perturbation theory in $t$, and is valid at all energy scales. One such construction was suggested in \cite{Dubovsky:2017cnj,Dubovsky:2018bmo}, as a theory of Jackiw-Teitelboim gravity     coupled to matter. A priori, this definition makes sense for either sign of $t$. It would be good to understand if this is indeed the case and, if this definition makes sense for $t<0$, which UV completion it corresponds to from our perspective.

Similarly, holography could provide an independent definition of these theories, at least in a large $c$ expansion. The current constructions via a cutoff in the bulk are not directly related to the $T\bar T$ deformations, despite many similarities. It would be interesting to find some consistent holographic dual for them (for $t<0$ this would involve a specific UV completion).

The construction of this paper can be generalized to other, related, theories. An example is a $J\bar T$ deformed CFT, which was discussed recently in \cite{Guica:2017lia,Bzowski:2018pcy,Chakraborty:2018vja,Apolo:2018qpq}. Using our techniques, one can show that the torus partition sum and spectrum of this theory can be constructed starting from modular covariance and the assumption that the spectrum of energies and charges of the deformed theory depends only on those of the undeformed theory \cite{toappear}.

If the original CFT has left and right moving currents $J$, $\bar J$, one can further generalize the discussion to general perturbations of the form $J\bar J$, $J\bar T$, $T\bar J$, $T\bar T$ and linear combinations thereof. Our construction may be useful for studying the resulting theories.
\section*{Acknowledgements}

The work of OA and AG was supported in part  by the I-CORE program of the Planning and Budgeting Committee and the Israel Science Foundation (grant number 1937/12) and by an Israel Science Foundation center for excellence grant (grant number 1989/14). The work of OA was also supported by the Minerva foundation with funding from the Federal German Ministry for Education and Research. OA is the Samuel Sebba Professorial Chair of Pure and Applied Physics. The work of SD and YJ is supported by the NCCR SwissMAP, funded by the Swiss National Science Foundation. 
The work of DK is supported in part by DOE grant DE-SC0009924. DK thanks the Hebrew University, Tel Aviv University and  the Weizmann Institute for hospitality during part of this work.


\par \vspace*{.5cm}\par

%

\begingroup\begin{thebibliography}{10}
	
	\bibitem{Zamolodchikov:2004ce}
	A.~B. Zamolodchikov, {\it {Expectation value of composite field T anti-T in
			two-dimensional quantum field theory}},
	\href{http://arxiv.org/abs/hep-th/0401146}{{\ttfamily arXiv:hep-th/0401146
			[hep-th]}}.
	
	\bibitem{Smirnov:2016lqw}
	F.~A. Smirnov and A.~B. Zamolodchikov, {\it {On space of integrable quantum
			field theories}},
	\href{http://dx.doi.org/10.1016/j.nuclphysb.2016.12.014}{{\sf Nucl. Phys.}
		{\sf {B915} }{\sf (2017) }{\sf 363--383}},
	\href{http://arxiv.org/abs/1608.05499}{{\ttfamily arXiv:1608.05499 [hep-th]}}.
	
	\bibitem{Cavaglia:2016oda}
	A.~Cavagli{\`a}, S.~Negro, I.~M. Sz{\'e}cs{\'e}nyi, and R.~Tateo, {\it {$T
			\bar{T}$-deformed 2D Quantum Field Theories}},
	\href{http://dx.doi.org/10.1007/JHEP10(2016)112}{{\sf JHEP} {\sf {10} }{\sf
			(2016) }{\sf 112}},
	\href{http://arxiv.org/abs/1608.05534}{{\ttfamily arXiv:1608.05534 [hep-th]}}.
	
	\bibitem{McGough:2016lol}
	L.~McGough, M.~Mezei, and H.~Verlinde, {\it {Moving the CFT into the bulk with
			$ T\overline{T} $}},  \href{http://dx.doi.org/10.1007/JHEP04(2018)010}{{\sf
			JHEP} {\sf {04} }{\sf (2018) }{\sf 010}},
	\href{http://arxiv.org/abs/1611.03470}{{\ttfamily arXiv:1611.03470 [hep-th]}}.
	
	\bibitem{Giveon:2017nie}
	A.~Giveon, N.~Itzhaki, and D.~Kutasov, {\it {$ \mathrm{T}\overline{\mathrm{T}}
			$ and LST}},  \href{http://dx.doi.org/10.1007/JHEP07(2017)122}{{\sf JHEP}
		{\sf {07} }{\sf (2017) }{\sf 122}},
	\href{http://arxiv.org/abs/1701.05576}{{\ttfamily arXiv:1701.05576 [hep-th]}}.
	
	\bibitem{Dubovsky:2017cnj}
	S.~Dubovsky, V.~Gorbenko, and M.~Mirbabayi, {\it {Asymptotic fragility, near
			AdS$_{2}$ holography and $ T\overline{T} $}},
	\href{http://dx.doi.org/10.1007/JHEP09(2017)136}{{\sf JHEP} {\sf {09} }{\sf
			(2017) }{\sf 136}},
	\href{http://arxiv.org/abs/1706.06604}{{\ttfamily arXiv:1706.06604 [hep-th]}}.
	
	\bibitem{Giveon:2017myj}
	A.~Giveon, N.~Itzhaki, and D.~Kutasov, {\it {A solvable irrelevant deformation
			of AdS$_{3}$/CFT$_{2}$}},
	\href{http://dx.doi.org/10.1007/JHEP12(2017)155}{{\sf JHEP} {\sf {12} }{\sf
			(2017) }{\sf 155}},
	\href{http://arxiv.org/abs/1707.05800}{{\ttfamily arXiv:1707.05800 [hep-th]}}.
	
	\bibitem{Shyam:2017znq}
	V.~Shyam, {\it {Background independent holographic dual to $T\bar{T}$ deformed
			CFT with large central charge in 2 dimensions}},
	\href{http://dx.doi.org/10.1007/JHEP10(2017)108}{{\sf JHEP} {\sf {10} }{\sf
			(2017) }{\sf 108}},
	\href{http://arxiv.org/abs/1707.08118}{{\ttfamily arXiv:1707.08118 [hep-th]}}.
	
	\bibitem{Asrat:2017tzd}
	M.~Asrat, A.~Giveon, N.~Itzhaki, and D.~Kutasov, {\it {Holography Beyond AdS}},
	\href{http://arxiv.org/abs/1711.02690}{{\ttfamily arXiv:1711.02690 [hep-th]}}.
	
	\bibitem{Giribet:2017imm}
	G.~Giribet, {\it {$T\bar{T}$-deformations, AdS/CFT and correlation functions}},
	\href{http://dx.doi.org/10.1007/JHEP02(2018)114}{{\sf JHEP} {\sf {02} }{\sf
			(2018) }{\sf 114}},
	\href{http://arxiv.org/abs/1711.02716}{{\ttfamily arXiv:1711.02716 [hep-th]}}.
	
	\bibitem{Kraus:2018xrn}
	P.~Kraus, J.~Liu, and D.~Marolf, {\it {Cutoff AdS$_{3}$ versus the $
			T\overline{T} $ deformation}},
	\href{http://dx.doi.org/10.1007/JHEP07(2018)027}{{\sf JHEP} {\sf {07} }{\sf
			(2018) }{\sf 027}},
	\href{http://arxiv.org/abs/1801.02714}{{\ttfamily arXiv:1801.02714 [hep-th]}}.
	
	\bibitem{Cardy:2018sdv}
	J.~Cardy, {\it {The $T\overline T$ deformation of quantum field theory as a
			stochastic process}},
	\href{http://arxiv.org/abs/1801.06895}{{\ttfamily arXiv:1801.06895 [hep-th]}}.
	
	\bibitem{Cottrell:2018skz}
	W.~Cottrell and A.~Hashimoto, {\it {Comments on $T \bar T$ double trace
			deformations and boundary conditions}},
	\href{http://arxiv.org/abs/1801.09708}{{\ttfamily arXiv:1801.09708 [hep-th]}}.
	
	\bibitem{Aharony:2018vux}
	O.~Aharony and T.~Vaknin, {\it {The TT* deformation at large central charge}},
	\href{http://arxiv.org/abs/1803.00100}{{\ttfamily arXiv:1803.00100 [hep-th]}}.
	
	\bibitem{Dubovsky:2018bmo}
	S.~Dubovsky, V.~Gorbenko, and G.~Hern{\'a}ndez-Chifflet, {\it {$T\bar{T}$
			Partition Function from Topological Gravity}},
	\href{http://arxiv.org/abs/1805.07386}{{\ttfamily arXiv:1805.07386 [hep-th]}}.
	
	\bibitem{Taylor:2018xcy}
	M.~Taylor, {\it {TT deformations in general dimensions}},
	\href{http://arxiv.org/abs/1805.10287}{{\ttfamily arXiv:1805.10287 [hep-th]}}.
	
	\bibitem{Bonelli:2018kik}
	G.~Bonelli, N.~Doroud, and M.~Zhu, {\it {$T\bar T$-deformations in closed
			form}},
	\href{http://arxiv.org/abs/1804.10967}{{\ttfamily arXiv:1804.10967 [hep-th]}}.
	
	\bibitem{Datta:2018thy}
	S.~Datta and Y.~Jiang, {\it {$T\bar{T}$ deformed partition functions}},
	\href{http://arxiv.org/abs/1806.07426}{{\ttfamily arXiv:1806.07426 [hep-th]}}.
	
	\bibitem{Donnelly:2018bef}
	W.~Donnelly and V.~Shyam, {\it {Entanglement entropy and $T \overline{T}$
			deformation}},
	\href{http://arxiv.org/abs/1806.07444}{{\ttfamily arXiv:1806.07444 [hep-th]}}.
	
	\bibitem{Babaro:2018cmq}
	J.~P. Babaro, V.~F. Foit, G.~Giribet, and M.~Leoni, {\it {$T\bar{T}$ type
			deformation in the presence of a boundary}},
	\href{http://arxiv.org/abs/1806.10713}{{\ttfamily arXiv:1806.10713 [hep-th]}}.
	
	\bibitem{Conti:2018jho}
	R.~Conti, L.~Iannella, S.~Negro, and R.~Tateo, {\it {Generalised Born-Infeld
			models, Lax operators and the $\textrm{T} \bar{\textrm{T}}$ perturbation}},
	\href{http://arxiv.org/abs/1806.11515}{{\ttfamily arXiv:1806.11515 [hep-th]}}.
	
	\bibitem{Chakraborty:2018kpr}
	S.~Chakraborty, A.~Giveon, N.~Itzhaki, and D.~Kutasov, {\it {Entanglement
			Beyond $\rm AdS$}},
	\href{http://arxiv.org/abs/1805.06286}{{\ttfamily arXiv:1805.06286 [hep-th]}}.
	
	\bibitem{Chen:2018eqk}
	B.~Chen, L.~Chen, and P.-x. Hao, {\it {Entanglement Entropy in
			$T\overline{T}$-Deformed CFT}},
	\href{http://arxiv.org/abs/1807.08293}{{\ttfamily arXiv:1807.08293 [hep-th]}}.
	
	\bibitem{Zagier}
	J.~H. Bruinier, G.~van~der Geer, G.~Harder, and D.~Zagier, {\it The 1-2-3 of
		modular forms: lectures at a summer school in nordfjordeid, norway}, .
	\newblock Springer Science \& Business Media, 2008.
	
	\bibitem{Witten:2001ua}
	E.~Witten, {\it {Multitrace operators, boundary conditions, and AdS / CFT
			correspondence}},
	\href{http://arxiv.org/abs/hep-th/0112258}{{\ttfamily arXiv:hep-th/0112258
			[hep-th]}}.
	
	\bibitem{Berkooz:2002ug}
	M.~Berkooz, A.~Sever, and A.~Shomer, {\it {'Double trace' deformations,
			boundary conditions and space-time singularities}},
	\href{http://dx.doi.org/10.1088/1126-6708/2002/05/034}{{\sf JHEP} {\sf {05}
		}{\sf (2002) }{\sf 034}},
	\href{http://arxiv.org/abs/hep-th/0112264}{{\ttfamily arXiv:hep-th/0112264
			[hep-th]}}.
	
	\bibitem{Dijkgraaf:2000fq}
	R.~Dijkgraaf, J.~M. Maldacena, G.~W. Moore, and E.~P. Verlinde, {\it {A Black
			hole Farey tail}},
	\href{http://arxiv.org/abs/hep-th/0005003}{{\ttfamily arXiv:hep-th/0005003
			[hep-th]}}.
	
	\bibitem{Manschot:2007ha}
	J.~Manschot and G.~W. Moore, {\it {A Modern Farey Tail}},
	\href{http://dx.doi.org/10.4310/CNTP.2010.v4.n1.a3}{{\sf Commun. Num. Theor.
			Phys.} {\sf {4} }{\sf (2010) }{\sf 103--159}},
	\href{http://arxiv.org/abs/0712.0573}{{\ttfamily arXiv:0712.0573 [hep-th]}}.
	
	\bibitem{Kutasov:1999xu}
	D.~Kutasov and N.~Seiberg, {\it {More comments on string theory on AdS(3)}},
	\href{http://dx.doi.org/10.1088/1126-6708/1999/04/008}{{\sf JHEP} {\sf {04}
		}{\sf (1999) }{\sf 008}},
	\href{http://arxiv.org/abs/hep-th/9903219}{{\ttfamily arXiv:hep-th/9903219
			[hep-th]}}.
	
	\bibitem{Giveon:1999zm}
	A.~Giveon, D.~Kutasov, and O.~Pelc, {\it {Holography for noncritical
			superstrings}},  \href{http://dx.doi.org/10.1088/1126-6708/1999/10/035}{{\sf
			JHEP} {\sf {10} }{\sf (1999) }{\sf 035}},
	\href{http://arxiv.org/abs/hep-th/9907178}{{\ttfamily arXiv:hep-th/9907178
			[hep-th]}}.
	
	\bibitem{Forste:1994wp}
	S.~Forste, {\it {A Truly marginal deformation of SL(2, R) in a null
			direction}},  \href{http://dx.doi.org/10.1016/0370-2693(94)91340-4}{{\sf
			Phys. Lett.} {\sf {B338} }{\sf (1994) }{\sf 36--39}},
	\href{http://arxiv.org/abs/hep-th/9407198}{{\ttfamily arXiv:hep-th/9407198
			[hep-th]}}.
	
	\bibitem{Argurio:2000tb}
	R.~Argurio, A.~Giveon, and A.~Shomer, {\it {Superstrings on AdS(3) and
			symmetric products}},
	\href{http://dx.doi.org/10.1088/1126-6708/2000/12/003}{{\sf JHEP} {\sf {12}
		}{\sf (2000) }{\sf 003}},
	\href{http://arxiv.org/abs/hep-th/0009242}{{\ttfamily arXiv:hep-th/0009242
			[hep-th]}}.
	
	\bibitem{Giveon:2005mi}
	A.~Giveon, D.~Kutasov, E.~Rabinovici, and A.~Sever, {\it {Phases of quantum
			gravity in AdS(3) and linear dilaton backgrounds}},
	\href{http://dx.doi.org/10.1016/j.nuclphysb.2005.04.015}{{\sf Nucl. Phys.}
		{\sf {B719} }{\sf (2005) }{\sf 3--34}},
	\href{http://arxiv.org/abs/hep-th/0503121}{{\ttfamily arXiv:hep-th/0503121
			[hep-th]}}.
	
	\bibitem{Guica:2017lia}
	M.~Guica, {\it {An integrable Lorentz-breaking deformation of two-dimensional
			CFTs}},
	\href{http://arxiv.org/abs/1710.08415}{{\ttfamily arXiv:1710.08415 [hep-th]}}.
	
	\bibitem{Bzowski:2018pcy}
	A.~Bzowski and M.~Guica, {\it {The holographic interpretation of $J \bar
			T$-deformed CFTs}},
	\href{http://arxiv.org/abs/1803.09753}{{\ttfamily arXiv:1803.09753 [hep-th]}}.
	
	\bibitem{Chakraborty:2018vja}
	S.~Chakraborty, A.~Giveon, and D.~Kutasov, {\it {$J\bar{T}$ deformed $CFT_2$
			and String Theory}},
	\href{http://arxiv.org/abs/1806.09667}{{\ttfamily arXiv:1806.09667 [hep-th]}}.
	
	\bibitem{Apolo:2018qpq}
	L.~Apolo and W.~Song, {\it {Strings on warped AdS$_3$ via $T\bar{J}$
			deformations}},
	\href{http://arxiv.org/abs/1806.10127}{{\ttfamily arXiv:1806.10127 [hep-th]}}.
	
	\bibitem{toappear}
	O.~Aharony, S.~Datta, A.~Giveon, Y.~Jiang and D.~Kutasov,
	{\it { Modular covariance and uniqueness of $J\bar{T}$ deformed CFTs}},
	\href{http://dx.doi.org/10.1007/JHEP01(2019)085}{{\sf JHEP} {\sf {01}
		}{\sf (2019) }{\sf 085}},
	\href{http://arxiv.org/abs/hep-th/1808.08978}{{\ttfamily arXiv:1808.08978 
			[hep-th]}}.
\end{thebibliography}\endgroup

\providecommand{\href}[2]{#2}

\end{document}